# A PRELIMINARY CLASSIFICATION SCHEME FOR THE CENTRAL REGIONS OF LATE-TYPE GALAXIES

SIDNEY VAN DEN BERGH*

Dominion Astrophysical Observatory, National Research Council
5071 West Saanich Road, Victoria, B.C., V8X 4M6, Canada

Electronic Mail: vandenbergh@dao.nrc.ca

*Visiting Astronomer, Canada-France-Hawaii Telescope

## ABSTRACT

The large-scale prints in The Carnegie Atlas of Galaxies have been used to formulate a classification scheme for the central regions of late-type galaxies. Systems that exhibit small bright central bulges or disks (type CB) are found to be of earlier Hubble type and of higher luminosity than galaxies that do not contain nuclei (type NN). Galaxies containing nuclear bars, or exhibiting central regions that are resolved into individual stars and knots, and galaxies with semi-stellar nuclei, are seen to have characteristics that are intermediate between those of types CB and NN. The presence or absence of a nucleus appears to be a useful criterion for distinguishing between spiral galaxies and Magellanic irregulars.

1. INTRODUCTION

The Palomar Sky Survey was first published 40 years ago. It contained a very large and uniform database of rather small-scale galaxy images. Inspection of these photographs showed that both the degree to which arm structure was developed in spirals, and the mean surface brightness of irregular galaxies, correlated with luminosity (van den Bergh 1960a,b,c) The usefulness of the Palomar Sky Survey was, however, limited by the small scale of its images and by the fact that the central regions of many galaxies were "burned out" on the Survey prints. As a result, the characteristics of the nuclear regions of spirals could not be used as classification criteria. The recent publication of The Carnegie Atlas of Galaxies (Sandage & Bedke 1994), which contains large-scale images of the overwhelming majority of Shapley-Ames galaxies (Sandage & Tammann 1981), now makes it possible to classify significant numbers of galaxies on the basis of the characteristics of their central regions. For previous work on the structure and population content of the central regions of galaxies, the reader is referred to Morgan (1958) and Morgan & Osterbrock (1969).

2. A PRELIMINARY CLASSIFICATION SCHEME

In the present investigation, all images in Volume II (late-type galaxies) of The Carnegie Atlas of Galaxies were inspected in an effort to derive useful classification criteria. A total of 342 central regions of Shapley-Ames galaxies were classified and are listed in Table 1. Also given in this Table are the Hubble types of these galaxies taken from Sandage & Tammann (1981). In cases where these authors list two Hubble types, only the first one is given in the Table. Many images reproduced in the catalog of Sandage & Bedke could not be classified because (1) the central part of the galaxy was overexposed, (2) strong dust absorption made classification difficult or impossible, (3) the galaxy was peculiar because of recent tidal interactions, starbursts, etc., or (4) the galaxy did not fit in a natural way into the preliminary classification scheme that is proposed below. The following is a brief description of the adopted classification system: (Numbers in square brackets refer to the panel number in the Sandage & Bedke (1994) Atlas.)

NN  Galaxy image contains no nucleus. The type example is NGC 2366 [327].

N   Image contains a star-like nucleus. Good examples are NGC 991 [245], NGC 5949 [279], NGC 6207 [274] and NGC 6503 [288].

SSN Galaxy has a semi-stellar nucleus. Good examples of this type are NGC 300 [261] and NGC 7793 [321].

CB  The galaxy is centred on a small bright central bulge or disk. The type standard is NGC 3726 [181]. Other good examples are NGC 1300 [154], NGC 1433 [158], NGC 2712 [165], NGC 3338 [173], NGC 4999 [159] and NGC 7038 [175]. In some cases (e.g. the Seyfert 1 galaxy NGC 4051 [180]), a semi-stellar nucleus is known to be present, but is not visible in the burned-out bulge of the image published by Sandage & Bedke (1994). In other cases, (e.g. NGC 1097 [201] and NGC 2903 [226]) the bright central region appears to be produced by a disk of HII regions and young OB stars, rather than by a bulge consisting of old or intermediate-age stars. In galaxies of types SBb and SBc, the central bulge may have a non-circular outline.

NB  In galaxies such as NGC 5112 [248], a nuclear bar-like structure is present in the galactic center. Other good examples of this type are NGC 672 [307], NGC 4116 [306] and NGC 5669 [2990].

Tr  These are transitional objects that appear intermediate between spirals that have central bulges and objects having central regions that are resolved into stars and knots. Good examples are NGC 1313 [309] and NGC 4647 [278].

In the next section, some correlations between these classification types and other parameters will be examined.

## 3. LUMINOSITY DEPENDENCE OF CLASSIFICATION TYPES

Figure 1 shows a plot of galaxy magnitude MB [this is of Sandage & Tammann (1981), with Ho = 50 kms-1 Mpc-1 adopted for most distant galaxies] versus Hubble type for galaxies classified as CB. These objects, which have bright central bulges or disks, are seen to be strongly concentrated in the region with Hubble types Sb-Sc and MB < -20. [Galaxies with less certain classification types CB: are observed to have a slightly larger scatter in the magnitude versus Hubble type diagram.]

Figure 2 shows an absolute magnitude versus Hubble type plot for those galaxies in Volume II of the Sandage & Bedke (1994) Atlas which do not appear to have nuclei (types NN and NN:). These objects mostly have MB > -20 and are mainly of Hubble-de Vaucouleurs types in the range Sc-Sd-Im. Intercomparison of Fig. 1 and Fig. 2 shows that galaxies of types CB and NN occupy complimentary regions in the absolute magnitude versus Hubble type diagram. Galaxies without nuclei (type NN) are seen to have lower luminosities and later Hubble types than do galaxies with bright central bulges (type CB). Galaxies classified as being transitional (type Tr), and those with nuclear bars (type NB), have a distribution in MB versus Hubble type that is intermediate between those of CB galaxies on the one hand and objects of type NN on the other.

The difference between the luminosity distributions of galaxies classified CB and CB: and for galaxies of type NN and NN: in Volume II of The Carnegie Atlas of Galaxies is shown in Fig. 3. This Figure shows that CB galaxies, which have central bulges (and presumably nuclei embedded within them) are more luminous than NN galaxies which do not contain nuclei. The only two NN galaxies in the present sample that are more luminous than MB = -19.5 are NGC 4945 [285] in the Centaurus cluster (which is one of the most peculiar galaxies in the sky) and NGC 5490 [288].

## 4. NUCLEI AND GALAXY LUMINOSITY

Among spheroidal galaxies, the fraction of all objects that contains a nucleus increases dramatically towards higher luminosity (van den Bergh 1986). A similar relationship also appears to hold for disk galaxies. Fig. 4 shows a plot of the frequency distribution of irregular and spiral galaxies in a volume-limited sample of nearby galaxies compiled by Kraan-Korteweg & Tammann (1979). [For objects beyond the Local Group, the Sculptor Group and the M81 Group, their distances are based on the Ho = 50 km s-1 Mpc-1]. The Figure shows that spirals (which have nuclei) dominate among luminous galaxies with MB < -16, whereas irregulars (which do not contain nuclei) are most common among disk galaxies with MB > -16. The tendency for the brightest galaxies to be nucleated therefore appears to hold for both disk and spheroidal galaxies.

Among nearby disk galaxies, M33 (BT = -19.1), NGC 7793 (MB = -18.8) and NGC 300 (BT = -18.6) have semi-stellar nuclei, whereas the LMC (BT = -18.4), the SMC (BT = -17.0) and NGC 6822 (BT = -15.2) do not. This suggests a transition at BT ~ -18.5 between disk systems that do, and that do not contain nuclei. The fact that NGC 205 (BT = -15.7) and M32 (BT = -15.5) do have nuclei indicates that the transition between spheroidal galaxies with and without nuclei may, on average, take place at a fainter luminosity in ellipticals than it does in disk galaxies.

There are two well-known galaxies that appear to provide counter examples to the notion that spirals contain nuclei but that irregulars do not. These are the Large Magellanic Cloud and NGC 4449. The classification of these two systems will be discussed in more detail below.

### 4.1 The Large Magellanic Cloud

The idea that the LMC is a barred spiral was introduced by de Vaucouleurs (1954). Subsequently, de Vaucouleurs & Freeman (1972) showed that the long "spiral arm" that provided the strongest support for the SBm classification of the LMC was, in fact, a Galactic foreground feature. The classification of the Large Cloud as an irregular would be consistent with the observation that this object does not contain a nucleus. It is of interest to note that Magellanic irregular galaxies exhibit the same dichotomy between normal and barred objects that is encountered among spirals. The LMC is, perhaps, the best-known example of a barred irregular, whereas the SMC is a normal irregular. Since S0, spiral and irregular galaxies may occur as both normal and as barred objects, one should probably regard bar formation as a "flavor" that can occur among all disk galaxies. Among the relatively nearby galaxies listed in the Kraan-Korteweg & Tammann (1979) catalog, there is no significant difference between the luminosity distributions of barred and of unbarred disk galaxies.

### 4.2 NGC 4449

Hubble (1926, 1936) defined his morphological classification system for galaxies in terms of giant or supergiant type examples. In particular, he used the luminous object NGC 4449 [326] as the type-example for irregular galaxies. In some ways, this choice of proto-type may have been unfortunate because NGC 4449, though lacking rotational symmetry, does appear to contain a well-developed (although not dominant) nucleus (see Fig. 5).  It has become a source of some confusion that NGC 4449 was classified as Ir by Hubble (1936) in The Realm of the Nebulae and by Sandage (1961) in The Hubble Atlas of Galaxies, but as Sm by Sandage & Tammann (1981) in A Revised Shapley-Ames Catalog of Bright Galaxies.  In fact, there appears to be a systematic deviation between the classification types of late-type galaxies assigned by Sandage & Tammann (1981) and those by other authors.  Of the 17 Northern Shapley-Ames galaxies which van den Bergh (1960c) assigns to type Ir, only one (6%) are classified as an irregular by Sandage & Tammann (1981).  These authors classify the remaining 16 objects as spirals.  By the same token, only one (9%) of the 11 galaxies called Ir by Humason, Mayall & Sandage (1956) are classified as irregular by Sandage & Tammann (1981).  However, if NGC 4449 is classified as a Magellanic irregular galaxy, then the apparent presence of a nucleus is an anomaly.  Possibly, the "nucleus" of this object is, in fact, a more-or-less centrally located enormous HII region and star forming complex similar to that which is observed in the type NN galaxy NGC 4861 [327].  Clearly, it would be very interesting to test this hypothesis by making radial velocity studies of the central region of NGC 4449. Such observations could establish if the bright star forming complex in the galaxy is, or is not, its dynamical nucleus.  The referee of this paper (Jay Gallagher) has emphasized the fact that some galaxies are known to have off-center bars and that some galactic nuclei might also be off-center.

5.   SUMMARY AND CONCLUSION
    Classifications have been made of 345 late-type galaxies in the Carnegie Atlas of Galaxies.  Galaxies of type CB (which have small bright nuclear bulges or bright centrally located disks) are found to be both more luminous, and of earlier type, than are galaxies of type NN (which do not contain nuclei).  It is suggested that the presence or absence of a nucleus in a late-type galaxy may be used as a criterion to distinguish in an objective fashion between spiral and irregular galaxies. It is also pointed out that galaxies of types (S0, Spiral, Ir) can occur in a normal or in a barred "flavor".  The transition between systems with, and without, nuclei may occur at a fainter luminosity level for ellipticals than it does for disk galaxies.

     I thank Chris Pritchet for providing me with a tape of our CFHT image of NGC 4449 and David Duncan for his help in producing Fig. 1.  I am also indebted to Jay Gallagher for discussions about NGC 4449 and its nucleus, to Gerard de Vaucouleurs for references to early classifications of the LMC, and to Janet Currie for typing the manuscript.

Table 1 – Classifications of late-type galaxies

| Galaxy | Hubble | Type | Galaxy | Hubble | Type | Galaxy | Hubble | Type | Galaxy | Hubble | Type |
|---|---|---|---|---|---|---|---|---|---|---|---|
| N24    | Sc   | Tr   | N895  | Sc   | CB   | N1512 | SBb  | CB   | N2441 | Sc   | CB   |
| N45    | Scd  | CB:  | N925  | SBc  | NB:  | N1518 | Sc   | NB   | N2500 | Sc   | NB?  |
| N95    | Sc   | CB   | N941  | Scd  | Tr   | N1536 | SBc  | NB   | N2525 | SBc  | CB   |
| N151   | SBbc | CB   | N958  | Sbc  | CB   | I2056 | Sc   | N    | N2523 | SBb  | CB:  |
| N157   | Sc   | SSN: | N976  | Sbc  | CB:  | N1559 | SBc  | NB   | N2537 | Sc   | NN   |
| N255   | SBc  | CB   | N991  | Sc   | N    | N1617 | Sa   | CB   | N2545 | SBc  | CB   |
| N247   | Sc   | Tr   | N1035 | Sc:  | Tr:  | N1659 | Sc   | SSN  | N2552 | Sc   | NN   |
| N255   | SBc  | CB   | N1042 | Sc   | SSN  | N1688 | SBc  | NB   | N2608 | Sbc  | SSN: |
| SMC    | Im   | NN   | N1058 | Sc   | CB:  | N1744 | SBcd | NB   | N2642 | SBb  | CB   |
| N300   | Sc   | SSN  | N1073 | SBc  | NB:  | N1796 | SBc  | NN   | N2712 | SBb  | CB   |
| N309   | Sc   | CB   | N1079 | Sa   | CB:  | N1784 | SBbc | CB:  | N2742 | Sc   | CB   |
| New 1  | SBc  | CB:  | N1084 | Sc   | SSN: | N1792 | Sc   | SSN  | N2763 | Sc   | CB   |
| N406   | Sc   | CB:  | N1090 | SBc  | CB   | HA85-1| Sc   | CB   | N2776 | Sc   | CB   |
| N450   | Sc   | CB   | N1097 | SBbc | CB   | LMC   | SBm  | NN   | N2748 | Sc   | Tr   |
| N470   | Sbc  | SSN  | N1156 | Sm   | NN   | N2082 | Sc   | Tr   | N2835 | SBc  | SSN: |
| N514   | Sc   | CB   | N1232 | Sc   | CB   | N2188 | Scd  | NN   | N2903 | Sc   | CB:  |
| N521   | SBc  | CB   | N1241 | SBbc | CB   | N2207 | Sc   | CB   | N2942 | Sc   | SSN  |
| N578   | Sc   | CB:  | N1300 | SBb  | CB   | N2223 | SBbc | CB   | N2907 | S0   | CB:  |
| N625   | Am   | NN   | N1313 | SBc  | Tr   | N2339 | SBc  | SSN  | N2976 | Sd   | NN   |
| N628   | Sc   | CB   | N1359 | Sc   | NB   | N2276 | Sc   | SSN  | N2998 | Sc   | CB:  |
| N672   | SBc  | NB   | N1376 | Sc   | CB:  | N2336 | SBbc | CB   | N3003 | Sc:  | NB?  |
| N685   | SBc  | CB:  | N1433 | SBb  | CB   | N2397 | Sc   | N?   | N3059 | SBc  | NB:  |
| N782   | SBb  | CB   | N1437 | Sc   | CB   | N2366 | SBm  | NN   | N3041 | Am   | CB   |
| I1783  | Sbc  | CB:  | N1493 | SBc  | NB   | N2427 | Sc   | NB?  | N3052 | Sc   | SSN  |
| N864   | Sbc  | SSN  | N1494 | Scd  | Tr   | N2442 | SBbc | CB   | N3054 | SBbc | CB:  |

| Galaxy | Hubble | Type | Galaxy | Hubble | Type | Galaxy | Hubble | Type | Galaxy | Hubble | Type |
|---|---|---|---|---|---|---|---|---|---|---|---|
| N3055 | Sc   | NB:  | N3486 | Sc   | CB:  | N3938 | Sc   | CB   | N4234 | SBc  | NB   |
| N3109 | Sm   | NN   | N3495 | Sc   | Tr   | N3949 | Sc   | CB   | N4237 | Sc   | CB   |
| I2537 | Sc   | CB:  | N3511 | Sc   | N:   | N3953 | SBbc | CB   | N4242 | SBd  | SSN  |
| N3124 | SBbc | CB   | N3510 | SBc  | NB   | N3956 | Sc   | NB?  | N4294 | SBc  | SSN: |
| N3145 | SBbc | CB   | N3513 | SBc  | NB   | N3963 | SBc  | SSN  | N4299 | Sd   | NN   |
| N3184 | Sc   | CB   | I2627 | Sc   | CB   | N3992 | SBb  | CB:  | N4303 | Sc   | SSN  |
| N3200 | Sb   | CB   | N3549 | Sbc  | SSN  | I749  | SBc  | SSN  | N4321 | Sc   | CB   |
| N3259 | Sb   | SSN  | N3556 | Sc   | NN?  | N4041 | Sc   | CB   | I3253 | Sc   | CB:  |
| N3287 | SBbc | NB   | N3596 | Sc   | CB   | N4062 | Sc   | SSN  | N4389 | SB   | NB:  |
| N3294 | Sc   | CB   | N3614 | Sc   | CB   | N4085 | Sc   | Tr?  | N4385 | SBbc | CB   |
| N3318 | SBbc | SSN  | N3629 | Sc   | CB:  | N4088 | Sc   | SSN  | N4395 | Sd   | SSN  |
| N3319 | SBc  | NB   | N3646 | Sbc  | SSN  | I2995 | Sc   | Tr   | N4414 | Sc   | CB:  |
| N3338 | Sbc  | CB   | N3666 | Sc   | CB   | N4096 | Sc   | CB   | N4412 | SBbc | SSN: |
| N3344 | SBbc | CB   | N3664 | SBm  | NB:  | N4100 | Sc   | CB   | N4449 | Sm   | NN   |
| N3346 | SBc  | NB   | N3686 | SBbc | CB:  | N4116 | SBc  | NB   | N4535 | SBc  | CB   |
| N3351 | SBb  | CB   | N3687 | SBbc | CB:  | N4123 | SBbc | CB   | N4540 | Scd  | Tr   |
| N3367 | SBc  | CB   | N3691 | S    | NN   | N4136 | Sc   | CB:  | N4559 | Sc   | SSN  |
| N3389 | Sc   | CB:  | N3720 | Sbc  | SNN  | N4145 | SBc  | CB   | N4567 | Sc   | CB:  |
| N3423 | Sc   | CB:  | N3726 | Sc   | CB   | N4162 | Sc   | CB:  | N4571 | Sc   | CB   |
| N3430 | Sbc  | SSN  | N3732 | Sc   | N    | N4183 | Scd  | Tr   | N4580 | Sbc  | SSN  |
| N3433 | Sc   | CB   | N3738 | Sd   | NN   | N4190 | Sm   | NN   | N4592 | Scd  | Tr   |
| N3445 | Sc   | Tr   | N3735 | Sc   | CB   | N4189 | SBc  | CB   | N4593 | SBb  | CB   |
| N3464 | Sc   | CB   | N3780 | Sc   | CB   | N4212 | Sc   | CB   | N4595 | Sc   | SSN: |
| N3478 | Sc   | SSN  | N3782 | SBm  | NN   | N4219 | Sbc  | CB:  | N4596 | SBa  | NB   |
| N3485 | SBbc | CB   | N3877 | Sc   | SSN  | N4236 | SBd  | NN   | N4597 | SBc  | NN:  |

| Galaxy | Hubble | Type | Galaxy | Hubble | Type | Galaxy | Hubble | Type | Galaxy | Hubble | Type |
|---|---|---|---|---|---|---|---|---|---|---|---|
| N4602 | Sc | CB: | N5068 | SBc | NB | N5468 | Sc | CB | N5907 | Sc | CB |
| N4603 | Sc | CB | N5085 | Sc | CB | N5494 | Sc | CB | N5921 | SBbc | CB |
| N4632 | Sc | Tr | N5088 | Sc | Tr | N5530 | Sc | CB | N5949 | Sc | N |
| N4639 | SBb | CB: | N5112 | Sc | NB | N5556 | SBc | NB | N5936 | Sc | SSN: |
| N4647 | Sc | Tr | N5156 | SBbc | CB | N5585 | Sd | CB | N5985 | SBb | CB |
| N4653 | Sc | CB | N5161 | Sc | CB | N5584 | Sc | SSN: | N5984 | SBcd | NB: |
| N4656 | Im | Tr: | N5204 | Sd | Tr | N5597 | SBc | CB: | N5967 | Sc | CB |
| N4668 | SBc | NN: | N5236 | SBc | CB | N5605 | Sbc | CB | N6070 | Sc | CB |
| N4682 | Sc | SSN | N5247 | Sc | CB: | N5633 | Sbc | SSN | N6118 | Sc | CB |
| N4689 | Sc | CB: | N5297 | Sc | CB | N5653 | Sc | SSN | N6181 | Sc | CB: |
| New 3 | SBcd | Tr | N5301 | Sc | CB | N5660 | Sc | SSN | N6217 | SBbc | CB |
| N4712 | Sc | SSN | N5313 | S: | CB: | N5645 | Sc | NB | N6207 | Sc | N |
| N4731 | SBc | NB | N5324 | Sbc | SSN | N5643 | SBc | CB | N6239 | SBc | Tr |
| N4763 | SBbc | CB: | N5334 | SB: | Tr | N5669 | Sc | NB | N6412 | SBc | CB: |
| N4790 | Sd | Tr: | N5347 | SBb | CB | N5676 | Sc | CB | I4662 | Im | NN |
| New 4 | Sc | SSN | N5350 | SBbc | CB | N5690 | Sc | NN | N6503 | Sc | N |
| N4861 | SBm | NN | N5351 | SBb | SSN | N5728 | SBb | CB | N6574 | Sbc | SSN: |
| N4891 | SBbc | CB | N5362 | S | CB | N5756 | Sc | SSN | N6643 | Sc | SSN |
| N4928 | Sbc | SSN | N5376 | Sbc | CB | N5768 | Sc | SSN | I4710 | SBd | NN |
| N4939 | Sbc | CB | HA 72 | Sc | Tr | N5775 | Sc | Tr? | I4721 | Sc | CB |
| N4945 | Sc | NN | N5406 | Sc | CB: | N5792 | SBb | CB: | N6699 | Sbc | CB: |
| N4947 | Sbc | SSN | N5398 | SBc | SSN: | F 703 | Sc | CB | N6744 | Sbc | CB |
| N4981 | SBbc | CB | N5426 | Sbc | CB: | N5885 | SBc | CB | N6780 | Sbc | CB |
| N4999 | SBb | CB | N5427 | Sbc | CB: | N5899 | Sc | CB | N6808 | Sc | CB |
| N4995 | Sbc | CB: | N5457 | Sc | CB: | N5905 | SBb | CB | N6814 | Sbc | CB: |

| Galaxy | Hubble | Type | Galaxy | Hubble | Type | Galaxy | Hubble | Type |
|---|---|---|---|---|---|---|---|---|
| N6822 | Im | NN | N7171 | Sb | CB | N7462 | SBc | NB: |
| N6878 | Sc | CB | I5152 | Sdm | Tr | N7479 | SBbc | CB: |
| N6923 | SBbc | CB: | I5201 | SBcd | NB: | N7496 | SBc | CB |
| N6925 | Sbc | CB | N7300 | Sc | CB | N7541 | Sc | NB: |
| N6946 | Sc | CB | N7307 | SBc | Tr | N7640 | SBc | CB |
| N6951 | Sb | NB: | N7309 | Sc | CB | N7689 | Sc | SSN |
| I5039 | Sc | N? | N7314 | Sc | SSN | I5332 | Sc | CB |
| I5052 | Sd | NN | N7329 | SBcd | CB | N7721 | Sbc | SSN: |
| N6970 | Sc | N | N7361 | Sc | NN? | N7723 | SBb | CB |
| N6984 | SBbc | CB: | N7418 | Sc | CB: | N7741 | SBc | NB |
| N7038 | Sbc | CB | N7421 | SBcd | CB | N7755 | SBbc | CB |
| N7064 | Scd | NN | N7424 | Sc | NB | N7793 | Sd | SSN |
| N7070 | SBc | CB | I5273 | SBc | CB | | | |
| N7124 | Sbc | CB | N7448 | Sc | CB | | | |
| N7137 | Sc | SSN | N7456 | Sc | SSN? | | | |

# FIGURE LEGENDS

Fig. 1 Magnitude versus Hubble type diagram for late-type galaxies of type CB, which have bright central bulges. Most of these galaxies are seen to have Hubble types Sb-Sc and $M_B < -20$.

Fig. 2 Magnitude versus Hubble type diagram for late-type galaxies of type NN, which do not have nuclei. Most of these objects are seen to have Hubble types Sc-Sd-Sm-Im and $M_B > -20$.

Fig. 3 Luminosity distribution for CB galaxies (left) and for NN galaxies (right). The Figure shows that galaxies with small bright central bulges are more luminous than those that do not have nuclei.

Fig. 4 Luminosity distribution of spiral and irregular galaxies in the Kraan-Korteweg & Tammann (1979) catalog of nearby galaxies. Most spirals are seen to be brighter than $M_B = -17$, whereas the majority of irregulars are fainter than this limit.

Fig. 5 I-band CFHT exposure of NGC 4449 obtained by Pritchet and van den Bergh in 1984. The image was obtained with a 320 x 512 RCA chip having a scale of 0.41 arcsec/pixel. The nucleus is marked by an arrow. Note that this may be a rare example of an irregular galaxy that appears to contain a nucleus.